\documentclass[twoside,a4paper]{article}
\usepackage[latin1]{inputenc}
\usepackage{natbib}
\usepackage[dvips]{graphicx} 
\usepackage{amssymb} 
\usepackage{amsmath} 
\usepackage{EMILpek}
\usepackage[english,swedish]{babel}
\title{Fluctuations in type IV pilus retraction} 
\author{Martin Lind\'en\footnote{Department of Physics, Royal Inst of Technology, AlbaNova, 10691 Stockholm, Sweden}
\footnote{corresponding author: e-mail linden@kth.se} 
\and Emil Johansson$^*$
\and Ann-Beth Jonsson\footnote{Department of Medical
Biochemistry and Microbiology, Uppsala Biomedical Center, \mbox{Uppsala
University}, Box 582, 75123 Uppsala, Sweden}
\and Mats Wallin$^*$} 
\date{}
\begin{document}
\selectlanguage{english}
\bibliographystyle{pnas}
\maketitle
\section*{Abstract}\addcontentsline{toc}{section}{\bf Abstract}
The type IV pilus retraction motor is found in many important
bacterial pathogens. It is the strongest known linear motor protein
and is required for bacterial infectivity. We characterize the
dynamics of type IV pilus retraction in terms of a stochastic chemical
reaction model. We find that a two state model can describe the
experimental force velocity relation and qualitative dependence of ATP
concentration. The results indicate that the dynamics is limited by an
ATP-dependent step at low load and a force-dependent step at high
load, and that at least one step is effectively irreversible in the
measured range of forces.  The irreversible nature of the sub-step(s)
lead to interesting predictions for future experiments: We find
different parameterizations with mathematically identical force
velocity relations but different fluctuations (diffusion constant). We
also find a longer elementary step compared to an earlier analysis,
which agrees better with known facts about the structure of the pilus
filament and energetic considerations. We conclude that more
experimental data is needed, and that further retraction experiments
are likely to resolve interesting details and give valuable insights
into the PilT machinery. In light of our findings, the fluctuations of
the retraction dynamics emerge as a key property to be studied in
future experiments.
\section*{Introduction}\addcontentsline{toc}{section}{\bf Introduction}
The type IV pilus (TFP) retraction motor is interesting and important
for several reasons. Not only is it necessary for the infectivity of
many human pathogens \cite{mattick}, it is also the strongest known
linear motor protein, making it an interesting test ground for
understanding generation of large forces in nanosystems. The physics
of pilus retraction has recently been studied experimentally
\cite{merz00,maier02,maier04}, and it is of interest for theoretical
study, which we present here.

Physical modeling of motor proteins has seen a rapid development
recently, made possible by improved experimental techniques that
enable measurement on the single molecule level. It is possible to
apply general thermodynamic considerations to construct models which
do not require fully detailed knowledge of the molecular details of
the system. This can lead to new insights about the molecular
mechanism, make predictions that can be tested experimentally, and
suggest new experiments.

We will now briefly review the relevant facts known about the PilT
system. After that, we introduce the model with emphasis on the
underlying assumptions and the interpretation of the model parameters.
TFP are surface filaments crucial for initial adherence of certain
gram-negative bacteria to target host cells, DNA uptake, cell
signaling, and bacterial motility. The pili filaments consist of
thousands of pilin subunits that polymerize to helical filaments with
outer diameter of about 6 nm, a 4 nm pitch and 5 subunits per turn
\cite{mattick,pilstruct}. Bacterial motility (twitching motility) is
propelled by repeated extension and retraction of TFP, by which the
bacterium can pull itself forward on surfaces like glass plates or
target host cells \cite{merz00}. During retraction, the filament is
depolymerized, and the pilin subunits are stored in the cell membrane
\cite{skerker}. TFP are expressed by a wide range of gram-negative
bacteria \cite{mattick} including pathogens such as
\bakterie{Neisseria gonorrhoeae} \cite{merz00}, \bakterie{Myxococcus
xanthus} \cite{sun} and \bakterie{Pseudomonas aeruginosa}
\cite{skerker}. The retraction process is believed to be mediated by a
protein called PilT, which is a hexameric \cite{forest} motor protein
in the AAA family of ATPases \cite{mattick}. Pilus retraction in
\bakterie{N. gonorrhoeae} generates forces well above $100$ pN
\cite{maier04,maier02}, making PilT the strongest known linear motor
protein. The large force combined with the hexameric structure of PilT
indicates that retraction of a single pilin subunit may involve
hydrolysis of several ATP molecules.

The physics of pilus retraction has so far been studied experimentally
in some detail \cite{merz00,maier02,maier04}. The experimental data on
retraction velocity has previously been analyzed theoretically using a
model with one chemical reaction step with an Arrhenius type force
dependence, which describes the retraction at high loads
\cite{maier04}. In this paper, we will extend the analysis to two
different reaction steps, which is sufficient describe the existing
data for loads. We will also make predictions about fluctuations in
the retraction process, which are experimentally accessible. This
turns out to be crucial as we consider models with constrained load
patterns.  We find two families of parameterizations that describe the
data fairly well, give exactly the same velocity, but predict
qualitatively different fluctuations.  The constrained models also
give a partial explanation for the surprisingly short elementary step
in the one state model: Adding an extra state to account for behavior
a low loads changes the interpretation of the data for high loads,
resulting in a longer elementary step.  Nevertheless, one would like
to account for an even longer elementary step, to make the description
compatible with the known structure of the pilus filament. This work
represents an important improvement in this respect. It also indicates
that the pilus retraction system hides more interesting dynamics, and
that fluctuations are the key property to gain further insights about
this remarkable motor.
\section*{Mechanochemical step model}\label{SecModel}\addcontentsline{toc}{section}{\bf Mechanochemical step model}
Discrete mechanochemical models are well established in the
description of motor proteins, and have been used successfully to
describe the motion of proteins that walk along structural filaments
\cite{boal,howard,svoboda94,fisher99,fisher00,fisher01,fisher03}.  The
main idea is to describe the motion in terms of stochastic transitions
between meta-stable states in the reaction cycle of the protein. The
starting point is random walk between nearest neighbor states on a
one-dimensional lattice. Each lattice point corresponds to a
meta-stable state in the reaction that drives the motion.

\begin{figure}
  \begin{center}
    \includegraphics[]{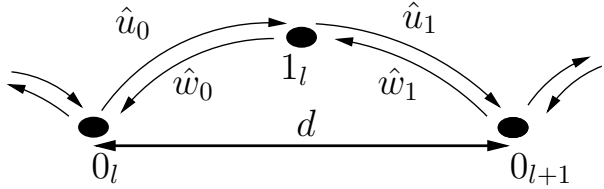}
  \end{center}
  \caption{$N=2$ mechanochemical step model.}\label{reaction}
\end{figure}

We consider an unbranched reaction with a period of $N$ steps, in
which each state is connected to two nearest neighbors. We denote
state $j$ in period $k$ by $j_k$, which corresponds to a position
$x(j_k)=x(j_0)+kd$ along the track, where $d$ is the spatial period of
the reaction (the elementary step).  The reaction is a Markov process
with non-negative forward and backward transition rates $\hat{u}_j$
and $\hat{w}_j$ respectively, as illustrated in figure \ref{reaction}
for the case $N=2$.  An exact solution due to Derrida \cite{derrida83}
gives analytic expressions for the steady state velocity $v$ and
diffusion constant $D$ for arbitrary transition rates and period $N$
\cite{fisher99,kolomeisky97}.

A model of this form can, at least in principle, be derived from a
theory of the microscopic degrees of freedom
\cite{bustamante00,reimann}, if we assume that the motor action
consists of fast reactions between meta-stable sub-states, which are
wells in a free energy landscape. Even so, the model does not have to
include all actual reactions; quickly decaying states can be lumped
together with slower, rate-limiting ones \cite{reimann}. An
alternative to the above mechanism of a fast ``working step'' that
produces work is the ``power stroke'', in which a fast reaction loads
potential energy to an internal degree of freedom, which in turn does
work through relaxation \cite{baker04}. Work following this approach
is under way.

Following Fisher and Kolomeisky \cite{fisher03,fisher01}, we take a
mechanochemical step model as a phenomenological starting point, and
assume that the chemical reactions obey an Arrhenius law
\cite{howard}. The non-negative transition rates depend on an opposing
load $F$, parallel to the track, and take the form \cite{howard}
\begin{equation}\label{load}\begin{array}{cc}
  \hat{u}_j=u_j e^{-Fc_j/\kBT},&
  \hat{w}_j=w_j e^{+Fd_j/\kBT}
  \end{array}\end{equation}
where $\kBT$ is Boltzmann's constant times temperature, $u_i$, $w_j$
are force independent rate constants, and $c_j,d_j$ are forward and
backward load distribution lengths.  $c_j,d_j$ can be interpreted as
positions of the activation barriers projected along the track.The
load lengths are expected to obey $\sum_{j=1}^N(d_j+c_j)=d$, where $d$
is the spatial period of the motion (the elementary step), which we
will discuss below. One can identify sub-steps of size $(d_j+c_j)$
between $j_k$ and $(j+1)_k$ \cite{fisher01,fisher03}.  

The simple form of the reaction rates and the correspondence between
load lengths and actual distances is not universally valid. It rests
on assumptions that the free energy wells corresponding to the
sub-states are narrow and similar in shape, and that the force is not
too large to make thermal fluctuations unimportant in the reaction
process \cite{bustamante00}.

In the following, we will consider $N=2$. There is not enough
experimental data to fit parameters for higher order models, and an
$N=1$ model can only describe the behavior in a limited range of
forces \cite{maier04}. Retraction experiments suggest a process with an
ATP-limited step at low loads and a force-limited step at high
loads \cite{maier02}, which suggests that a two-state model is
necessary to describe the full dynamics.  The steady state velocity 
given by \cite{derrida83,kolomeisky97,fisher99}
\begin{equation}\label{vN2}
v\equiv\lim_{t\to\infty}\frac{\mean{x(t)}}{t}=
\frac{(\hat{u}_0\hat{u}_1-\hat{w}_0\hat{w}_1)d}
{\hat{u}_0+\hat{u}_1+\hat{w}_1+\hat{w}_0}
\end{equation}
and the diffusion constant is 
\begin{equation}\label{DN2}
D=\frac{d^2}{2\sigma}(\hat{u}_0\hat{u}_1+\hat{w}_0\hat{w}_1)
-\frac{d^2}{\sigma^3}(\hat{u}_0\hat{u}_1-\hat{w}_0\hat{w}_1)^2
\end{equation}
The diffusion constant $D$ is defined as
$\big[\mean{x^2(t)}-\mean{x(t)}^2\big]/2t$ in the limit
$t\to\infty$. As is evident from the above equations, $v$ and $D$ do
not depend on which backward rate is associated with which forward
rate. Therefore it is not possible to uniquely identify sub-steps from
knowledge of $v$ and $D$ in a two state model.
\section*{Modeling retraction data}\label{sec:model}
\begin{figure}[t!]\begin{center}\rule{0pt}{1mm}\vspace{-1.5cm}\\
    \includegraphics{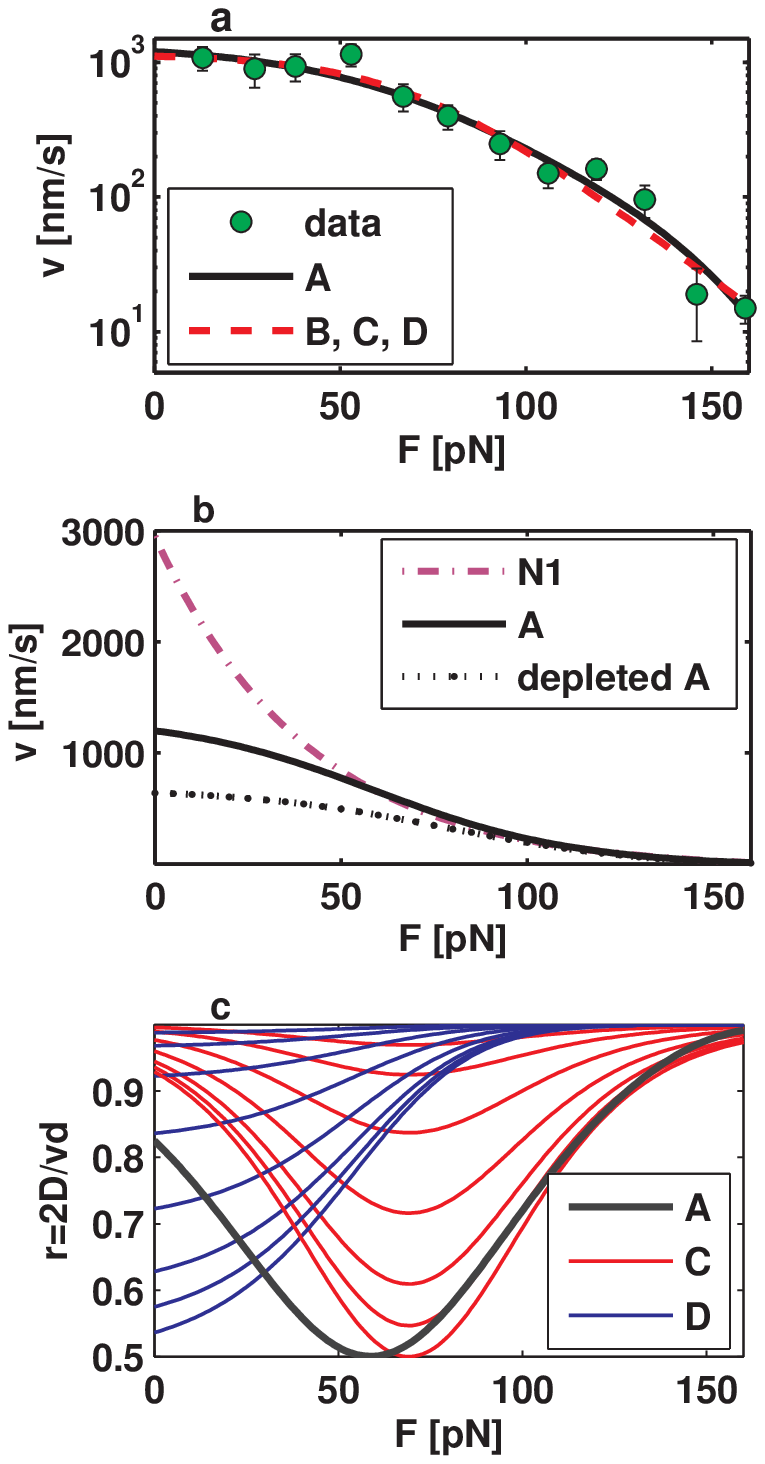}
    \caption{Force velocity relations for pilus retraction,
    parameterizations as defined in table \ref{MLtest}.  
    (a) Experimental retraction data for ``wild type''
    \bakterie{N. gonorrhoeae} \cite{maier04}, full $N=2$
    parameterizations \pars{A} (same as \pars{N2}), and restricted
    $N=2$ parameterizations \pars{B}, \pars{C} and \pars{D}. The
    experiment is well described by the $N=2$ model, and the result
    for \pars{A} and \pars{N2} is almost indistinguishable from
    \pars{B}, \pars{C} and \pars{D}.
    (b) High load analysis as in Ref.\ \cite{maier04} with an $N=1$
    model (\pars{N1}) compared to the \pars{A} parameterization. Also
    a model for an ATP-depleted version of \pars{A} by setting $u_0
    \to 0.5u_0$.  The ATP-depleted velocity tends to the non-depleted
    result (\pars{A}) at high loads, in qualitative agreement with
    experimental findings \cite{maier02}.  Also, the $N=1$ and $N=2$
    models agree at high load.
    (c) Predicted randomness $r=2D/vd$ for the different
    parameterizations \pars{A} (and \pars{N2}), \pars{C} and \pars{D}.
    Parameters \pars{B} is the limit $R_C\to0$ of \pars{C}, which is
    the lowest curve. The lowest curve of the \pars{D} parameter
    family is the limit $R_D\to4(u_0\ep{B})^2/u_1\ep{B}\approx175$
    s$^{-1}$.  
    The \pars{C} and \pars{D} families predict qualitatively different
    behavior, which should make it possible to distinguish
    experimentally between them even though they give exactly the same
    velocity. Also note that $0.5$ is a strict lower bound on randomness
    for models of this kind with $N=2$ \cite{koza02}.}\label{VFfit1}
\end{center}\end{figure}

\addcontentsline{toc}{section}{\bf Modeling retraction data}
We fitted Eqs.\ (\ref{load},\ref{vN2}) to retraction data for wild
type \bakterie{N. gonorrhoeae} at $330$ K from Ref.\ \cite{maier04},
with a maximum likelihood (ML) estimate. Assuming Gaussian errors,
this amounts to minimizing $-2\ln L=\sum_n (v(F_n)-v_n)^2/\sigma_n^2$
\cite{numrec}, where $F_n$, $v_n$ and $\sigma_n$ are experimental
forces, velocities and error bars respectively, and $L$ is the
(Gaussian) likelihood function.  The period $d$ is not an independent
parameter in the velocity, and was included in the transition
rates. As seen in Fig.\ \ref{VFfit1}, the data is well described. The
best fit parameters of the unrestricted $N=2$ model, which we denote
\pars{N2}, are shown in Tab.\ \ref{MLtest}, along with parameter sets
which are optimal when the parameter space is restricted in different
ways. We denote the restricted parameterizations \pars{A} and
\pars{B}, and also include \pars{N1}, the high force analysis of Ref.\
\cite{maier04}, which we reproduce for reference.

Some parameters in \pars{N2} are extremely small, and can be put to
zero without any significant difference in velocity, which is how the
\pars{A} parameterizations is constructed. This does not imply that
one reaction is completely irreversible, which would violate detailed
balance \cite{reimann}, but rather that one backward rate is small
enough to be negligible in the fit.\footnote{Interestingly, a discrete
$N=2$ model with one irreversible step can be derived as a limiting
case of certain class of simple ratchet models \cite{kolomeisky97}.}

There is no obvious relation between the load distribution lengths of
\pars{A} and \pars{N1}, which is surprising since \pars{N1} is a good
description of the high force regime. It is therefore tempting to ask
whether the force velocity relation can be equally well described by
other parameterizations, with at least one irreversible step and only
one force-dependent step. The minimal way to do this is to remove the
backward step completely from \pars{A}. This constrained
parameterization, \pars{B}, cannot be statistically rejected even at
80\% confidence. However, the rate-limiting step in \pars{B} is still
twice as large in \pars{N1}, a result which is retained even if only
high forces are considered (not shown). The reason for this difference
is the presence of a force independent step. To show this, we rewrite
the $N=1$ velocity in the form of an $N=2$ model with two equal
consecutive reactions. For simplicity, we neglect the weak force
dependence of the backward rate. With the notation $p\ep{X}$ for
parameter $p$ in parameterization $X$, we get:
\begin{equation}\label{N1toN2}
  v\ep{N1}=
  \frac{2d\ep{N1}\Big((u_0\ep{N1})^2e^{-2d_0F\ep{N1}/\kBT}-(w_0\ep{N1})^2\Big)}
       {2\big(u_0\ep{N1}e^{-Fd_0\ep{N1}/\kBT}+w_0\ep{N1}\big)}
\end{equation}
At forces where the numerator is dominating the force dependence, this
is similar to an $N=2$ model with one force independent step and twice
as long load distribution length on the other step. The factor 2 in
numerator and denominator shows the fact that the elementary step is
also twice as long, if the denominator is brought to the same form as
Eq.\ \eqref{vN2}, with four terms.  We conclude that the plateau in
$v(F)$ at low loads have implications for the interpretation of data
at high loads, and that the earlier estimate of the elementary
step in Ref.\ \cite{maier04} is too small by (at least) a factor 2. 

The \pars{B} parameterization has another interesting property. It has
exactly the same force velocity relation as two different parameter
families, \pars{C} and \pars{D}, which we now describe.  To construct
\pars{C} from \pars{B}, one adds an arbitrary force independent
backward reaction.  To keep the velocity unchanged, the force
dependent forward rate must be modified, but all other parameters in \pars{B} are kept.
\begin{equation}\label{CfromB}\begin{array}{l@{\,,\;}l@{\,,\;}l}
    w_1\ep{C}=R_C&d_1\ep{C}=0&
    u_1\ep{C}=u_1\ep{B}\Big(\frac{u_0\ep{B}+R_C}{u_0\ep{B}}\Big)
\end{array}\end{equation}
Here, the constant $R_C$ is non-negative but otherwise arbitrary, and
we retrieve the \pars{B} model in the limit $R_C\to 0$.  To construct
the \pars{D} parameter family from \pars{B}, we move the load
dependence from the forward step to a new (arbitrary) backward rate,
and modify both forward rates.
\begin{equation}\label{DfromB}
  \begin{array}{lcl}
    w_1\ep{D}=R_D,&R_D>(2u_0\ep{B})^2/u_1\ep{B},&d_1\ep{D}=c_1\ep{B}\\
    \multicolumn{2}{l@{\,,\;}}{u_{0,1}\ep{D}=\frac{u_1\ep{B}R_D}{2u_0\ep{B}}
    \bigg(1\pm\sqrt{1-\frac{(2u_0\ep{B})^2}{u_1\ep{B}R_D}}\;\bigg)}&
    c_1\ep{D}=0\end{array}
\end{equation}
$R_D$ again defines a parameter family, and the choice of sign in the
lower equation is irrelevant for $v$ and $D$.
\begin{table}[t!]\begin{center}
\begin{minipage}[t]{8.7cm}
  \begin{tabular}{|@{$\,$}c@{$\,$}|p{2.65cm}@{$\,$}|@{$\,$}c@{}|p{1cm}|l@{}|}
    \hline
    $X$ &comment&$k_X$&\mbox{-$2\ln L_\text{X}$}&parameters \\
    \hline\hline
    \pars{N2}& full $N=2$ model&               8&130.69&
    $\begin{array}{l@{=\,}l}
      du_0&1.35\e{3}\\ 
      c_0&3.6\e{-14}\\
      du_1&1.24\e{4}\\ 
      c_1&0.162\\
      dw_0&2.9\e{-14}\\
      d_0&4.4\e{-14}\\
      dw_1&14.3\e{-3}\\
      d_1&0.311\end{array}$\\
    \hline      
    \pars{A}& $w_0$, $c_0=0$& 5&130.69&
    \parbox[c]{2.5cm}{Same as N2 except $w_0$, $c_0=0$}\\
    \hline
    \pars{B}& $w_0$, $c_0$, $w_1=0$&3&132.14&
    $\begin{array}{l@{=\,}l}
      du_0&1.15\e{3}\\
      du_1&3.02\e{4}\\
      c_1&0.202\end{array}$\\
      \hline
      \pars{N1}&\parbox[c]{2.5cm}{$N$=1 model $F>60$ pN only}&4&-&
      $\begin{array}{l@{=\,}l}
	du_0&2.4\e{3}\\
	c_0&0.104\\
	dw_0&39\\ 
	d_0&3.2\e{-9}
      \end{array}$\\
      \hline
  \end{tabular}
\end{minipage}

  \caption{Maximum likelihood estimates of parameters. The products
    $du_i$, $dw_j$ are given in units of nm/s, and load lengths are in
    nm. $d$ is the (unknown) lattice period. $L_\text{X}$ is the
    maximum value of the likelihood function and $k_X$ the number of
    free parameters in parameterization $X$. If $X$ is constructed
    from parameterization $Y$ by constraining some parameters in $Y$,
    $X$ is rejected in favor of $Y$ if $2\ln(L_Y/L_X)$ is larger than
    the upper $\alpha$-percentile of a $\chi^2$ distribution with
    $(k_Y-k_X)$ degrees of freedom, where $1-\alpha$ is the chosen
    level of confidence \cite{NISTstatistics}.  \pars{A} cannot be
    rejected in favor of \pars{N2}; there is practically no
    difference. Even at only 80\% confidence, \pars{B} cannot be
    rejected in favor of \pars{N2} or \pars{A}. An unrestricted fit of
    the behavior at high load to an $N=1$ model gives the
    \pars{N1} parameters, reproducing the analysis of Ref.\
    \cite{maier04}.}\label{MLtest}
\end{center}
\end{table}

Using the obtained parameters, we can compute the diffusion constant
$D$. In the context of molecular motors, diffusion is commonly
analyzed in terms of the dimensionless randomness parameter $r=2D/vd$
\cite{fisher03,fisher01,svoboda94,fisher00,koza02,chen02}, which is
shown in Fig.\ \ref{VFfit1}c. The randomness is convenient in this
case, as it is independent of $d$, which is unknown for the PilT
motor. Although parameterizations \pars{B}, \pars{C} and \pars{D} give
equivalent force-velocity relations, they make different predictions
for the diffusion constant. We see that \pars{B} and \pars{C} make
qualitatively different predictions than \pars{D}, and that \pars{A}
is quite close to \pars{B} at high forces, but deviating at $F<60$
pN. We expect a measurement of the diffusion constant $D$ would
discriminate between the different parameterizations. \pars{N2} is
again indistinguishable from \pars{A} (not shown).

The velocity dependence on ATP concentration ([ATP]) has been studied
in experiments on ATP depleted bacteria, and two regimes where found
\cite{maier02}. At low loads, the velocity was strongly dependent on
[ATP], but at high loads, the velocity was the same for the depleted
strain as for the undepleted strain (``wild type''). A simple way to
include ATP dependence in a mechanochemical step model is to make one
forward rate (`binding reaction`) proportional to [ATP].  Figure
\ref{VFfit1}b shows the velocity relation for the \pars{A} parameters
with the load independent forward rate reduced by 50\%. We see that
the difference compared to the original \pars{A} parameters vanishes
for high load, in qualitative agreement with the experimental result
\cite{maier02}. Adding the [ATP] dependence to the other step gives
qualitatively different dependence (not shown). Also note that if
[ATP] dependence is added to the restricted parameterizations
\pars{B}-\pars{D} in this way, they still predict the same velocity
for all loads and [ATP].
\section*{Conclusion}\addcontentsline{toc}{section}{\bf Conclusion}

We interpret pilus retraction data on wild type
\bakterie{N. gonorrhoeae} in terms of a mechanochemical model, which
is a discrete random walk with steps between nearest
neighbors. Despite its simplicity, a description in these terms
contains interesting information about the free energy landscape of
the retraction reactions \cite{fisher03,fisher01,bustamante00}. 

We find that the experimental data for retraction velocity is well
described for all measured forces by several parameterizations, all of
which have one effectively force independent step and one irreversible
step. The model also captures the qualitative behavior of varying
[ATP], namely that the [ATP] dependence of velocity is strong at low
force and very weak at high force \cite{maier02}. As expected, the
binding reaction is not force dependent, but the unbinding might be.

We find several different parameterizations that give similar or
identical velocities, but make very different predictions for the
diffusion constant $D$ of the retraction. Without any assumptions
regarding the elementary step $d$, we predict the randomness $r=2D/dv$
as shown in Fig.\ \ref{VFfit1}c, which would be of considerable
interest to study experimentally. Such measurements may be able to
distinguish the different parameterizations
(\pars{A}, \pars{B}, \pars{C} and \pars{D}) and lead to additional
important insight into the pilus retraction mechanism.

We use a model that can account for the all measured forces, we make
predictions open to experimental test, and we get new result for the
elementary step $d$: $0.5$ nm for parameterization \pars{A} and $0.2$
nm for the \pars{B}, \pars{C} and \pars{D} parameters, compared to
$0.1$ nm for the $N=1$ model \cite{maier04}.  As we argued above, the
short step may be an artifact of using a model with too few states to
describe a restricted part of the force-velocity curve. Adding another
state to account for the behavior at low loads changes the
interpretation of the high load behavior.  

Even if we get at least twice as long step as in Ref.\ \cite{maier04},
it is still not obvious how our result fits with the known facts about
the system. Each pilin subunit contributes about $0.8$ nm to the
filament length \cite{maier04,maier02,mattick,pilstruct}, hence one
would like to account for at least one pilin subunit in a complete
description of a reaction cycle. In this light, our work is a great
improvement even if there are still a few Ångström left to account
for. One possibility is that a model with more states could account
for the missing length, just like going from one state to two states
did. Another is that the strong forces deform the sub-states and
destroy the correspondence between load lengths and physical
lengths \cite{bustamante00}. Since the existing data is well described
by our model, this question must probably be settled experimentally.
There is also a length associated with the involved energies and
forces: at the maximal measured force, $160$ pN, the free energy gain
from hydrolysis of one ATP molecule under physiological conditions
($\approx80$ pN nm \cite{boal,howard}) is enough to retract $0.5$ nm.
The evidence that retraction is powered by ATP
hydrolysis \cite{maier02} does not imply that it is powered only by ATP
hydrolysis. The above estimate indicates that more energy than that of
one ATP is needed to retract one subunit. Since PilT forms a hexamer,
it may hydrolyze up to six ATP in parallel, which would explain the
high force energetically. Another possibility is that free energy from
depolymerization of the filament is used in retraction.  It might be
possible to resolve details of for example a second binding event with
a model with more states.  However, models with more states have more
parameters, so more experimental data is needed to explore these
exciting possibilities.

This work extends previous physical modeling of the PilT system
\cite{maier04} in several ways. In particular, the fluctuations
(randomness parameter) emerge as a key property for further
theoretical and experimental study of the dynamics of pilus
retraction.

\section*{Acknowledgement}\addcontentsline{toc}{section}{\bf Acknowledgement}
This work was supported by the Swedish Research Council (MW 2003-5001,
ABJ 2001-6456, 2002-6240, 2004-4831), the Royal Institute
of Technology, the G\"oran Gustafsson Foundation, the Swedish Cancer
Society, and Uppsala University.
\addcontentsline{toc}{section}{\bf References}

\newpage

\newpage

\end{document}